# A RELIABLE SEMI-DISTRIBUTED LOAD BALANCING ARCHITECTURE OF HETEROGENEOUS WIRELESS NETWORKS

Md. Golam Rabiul Alam[1], Chayan Biswas[2], Naushin Nower[2], Mohammed Shafiul Alam Khan[2]

[1]Islamic University Chittagong, Chittagong-4203, Bangladesh
gra9710@yahoo.com
[2]Institute of Information Technology, University of Dhaka, Dhaka-1000 Bangladesh
chayan_ict@yahoo.com, naushin@iit.du.ac.bd, shafiul@univdhaka.edu

### *ABSTRACT*

*Now a day's Heterogeneous wireless network is a promising field of research interest. Various challenges exist in this hybrid combination like load balancing, resource management and so on. In this paper we introduce a reliable load balancing architecture for heterogeneous wireless communications to ensure certain level of quality of service. To conquer the problem of centralized and distributed design, a semi-distributed load balancing architecture for multiple access networks is introduced. In this grid based design multiple Load and Mobile Agent Management Units is incorporated. To prove the compactness of the design, integrated reliability, signalling overhead and total processing time is calculated. And finally simulation result shows that overall system performance is improved by enhancing reliability, reducing signalling overhead and processing time.*

## 1. INTRODUCTION

Today's wireless networks are highly heterogeneous with integration of multiple wireless network access interfaces. For the sake of anywhere-and-anytime connectivity in the next-generation networks, different 'wireless access' networks are integrated to form a heterogeneous network [1]. With the growing interest in various wireless networks, the need for dynamic resource utilization and efficient transmission of heterogeneous wireless networks is becoming increasingly important. Recent research focuses a variety of issues on heterogeneous wireless networks. Resource management, load balancing, reliability and QoS support are the most significant research issues on what we know as a heterogeneous system.

Load balancing is one of the key technologies in the convergence of heterogeneous wireless networks. It is a significant way to achieve the resource sharing over heterogeneous wireless networks, and it improves resource utilization, enlargement of system capacity as well as provision better services for users [2]. Generally load balancing depends on network architecture and the algorithm of load balancing. In this paper we provide efficient network architecture ideas so that the load of an entry system can be properly balanced. Network architecture can be classified as centralized, distributed, semi centralized and semi distributed [3], [4]. There are some problems in the first two mechanisms: the centralized one is relatively less reliable, while the distributed one has a large overhead [5].

To overcome the limitations of network architecture we have proposed a Semi-Distributed Architecture (SDA) for heterogeneous networks that efficiently reduces signal overhead, improves reliability and at the same time utilizes the system load by introducing a mobile agent. Mobile agents provide a support for dynamic load balancing since they can move across heterogeneous platforms and they carry application-specific code with them, rather than requiring pre-installation of that code on the destination machine [6]. Moreover, the agent-based paradigm also supports the disruptive nature of wireless links and alleviates its associated bandwidth limitations.

The rest of the paper is organized as follows, Section II reviews some of the related work found in the literature. The proposed design is elaborately depicted on Section III. The design is evaluated in Section IV. In Section V 'simulation result' is illustrated. Finally, we conclude this paper in Section VI with a discussion.

## 2. LITERATURE REVIEW

Lots of research was done regarding 'load balancing', especially in the area of 'wired homogeneous networks'. But 'load balancing' in wireless networks is the research issue of prominence as wireless communications are expanded day by day and multiple wireless networks work on the same grid.

In the reference [7], a dynamic load balancing architecture is proposed for heterogeneous wireless network environments. The authors of the paper have introduced the community concept, in which radio resources of various Radio Access Technologies (RATs) are managed together. They have also introduced the Community Resource Management (CRM) function, the Local Resource Management (LRM) function, and the Spectrum Manager (SM) in their architecture, where CRM builds and sends Community Resource Announcements (CRA) to Community Access Points (CAP). Here CRM enhances the utilization of radio resources and also reduces call blocking probability of individual RATs in the community. The main drawback of this method is the complexity of discovering suitable RATs. Discovering a RAT without knowing its operating information is a time-and-power-consuming task for Mobile Nodes (MNs). In this architecture, the performance improvement comes from a vertical handoff among RATs, but the cost of a vertical handoff is not studied. There is no backup- or failure-handling methodology existing in this architecture.

In reference [4], a semi-centralized-and-semi-distributed architecture (SCSDA) is proposed. In it, each base station of any access network type transfers load information to its six neighboring base stations. The architecture provides improvements by reducing blocking probability, dropping probability and thus advances overall system utilization. The major drawback of this architecture is the cost of the total number of handoffs. Both the periodic and non-periodic signaling overhead is huge.

Mobile agent-based load balancing architecture is introduced in reference [8]. A credit value is assigned with each of the agents. The associated credit depends on its affinity to a machine, current workload, communication behavior and mobility. The credit value of all agents are determined and then compared to balance loads and if any imbalance is found then an agent with a lower credit value is migrated to a relatively lightly loaded machine in the system. As a mobile agent is a program which can autonomously migrate from one host to another in a heterogeneous network, and can interact with other agents or resources [9], we have incorporated mobile agents to balance loads within a heterogeneous cell.

The authors of [10] considered general heterogeneous network architecture with two basic entities in the system, one is Mobile Nodes (MNs) and another one is Access Points (APs). They formulated the overhead of AP discovery which is divided into 'hello' messages and RREQ messages, and gives the simulation results. The proposed architecture has lack of reliability reflection.

In the paper [11], authors introduced Hierarchical Semi-Centralized Architecture (HSCA) for heterogeneous wireless networks. In this design, an Information Server (IS), a centralized server, exists at the top level of the hierarchy. And at the middle and lower levels of the hierarchy, RA (Resource Allotter) and RS (Resource Statistics) are introduced respectively, to manage the system load. The IS is responsible for maintaining backup of all RAs and also accountable for failure-handling of RAs. Since the IS maintains all information of RAs in the system, it creates a huge message-passing overhead. Again, to find a proper place for the IS from which all RAs can access it efficiently is quite impossible. Though there is a backup of the IS, it is not quite effective to organize a large network with a single IS.

## 3. DETAILS OF THE PROPOSED DESIGN

The details of the proposed architecture and also the signalling flow are discussed in the following two subsections.

### 3.1. Proposed Architecture

A distributed architecture for wireless networks is appealing due to the single-point failure of a centralized design. Our design is partial/semi-distributed in nature which is free from the huge overhead and delay problems of a fully distributed design. In our proposed semi-distributed architecture, a number of adjacent hexagonal cells create a basic grid. Since the system is heterogeneous, each cell may consist of different access networks. For simplicity, three types of access networks (UMTS, WLAN and WiMax) are shown in Figure 1. In each grid, a Load and Mobile Agent Management (LMMU) unit is introduced. LMMUs consist of three units; one unit is responsible for maintaining a resource inventory for each type of available access network in each cell, which is called a Resource Inventory (RI). Another unit, the Mobile Agent (MA) [9] in each cell is accountable for balancing load among various access networks within each cell.

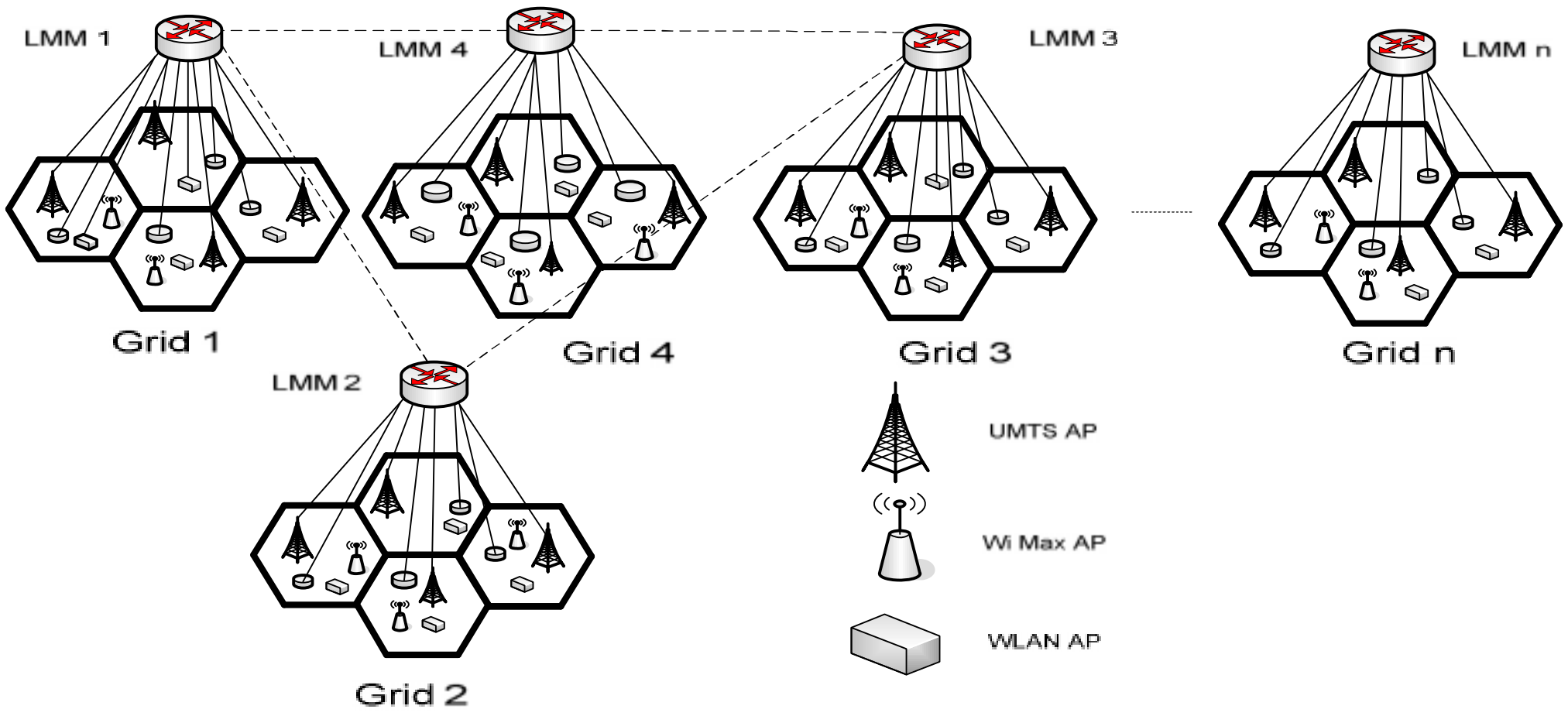

Figure 1. Semi distributed architecture

The third unit of LMMU is Load and Mobile Agent Manager (LMM), which is responsible for load balancing among adjacent cells in each grid and among adjacent grids.

To ensure reliability, each LMM is associated with its two neighboring LMMs so that if any of the LMMs goes down or is broken, then the backup LMM can take the responsibility for the failed one. In this proposed scheme if one LMM goes down, then only that network management is maintained by a backup LMM and the whole system, therefore, is not hampered.

In this research, we utilize the concept of a Bulletin Board (BB) [12], [13] for maintaining the repository of Load information. A Bulletin Board keeps information for a group of LMMs. To avoid a single point of failure, the replica of the BB is maintained, where the number of replica depends on the size of the network. When an LMM changes its state it just informs the BB. To balance load in the system each LMM exchanges information with BB. The design is called semi-distributed because the load distribution decision is made by consulting a group of LMMs through the BB. Therefore this architecture is free from a large signalling overhead.

## 3.2. Signalling Flow

Usually RI calculates the load and resources of its jurisdiction cell according to the wireless parameters received from the AP, and then transfers the load, resources and location information to the LMM. On the basis of load balancing algorithm, next the LMM transfers load balancing information to the RI. Then the RI changes load balancing information to load balancing instructions and transfers the load balancing instructions to the AP. The signalling flow chart is shown in Figure 2. Only when LMM changes its state (under loaded to balance and so on) it then informs BB. Again when a LMM searches for suitable LMM, it then asks BB for the better one.

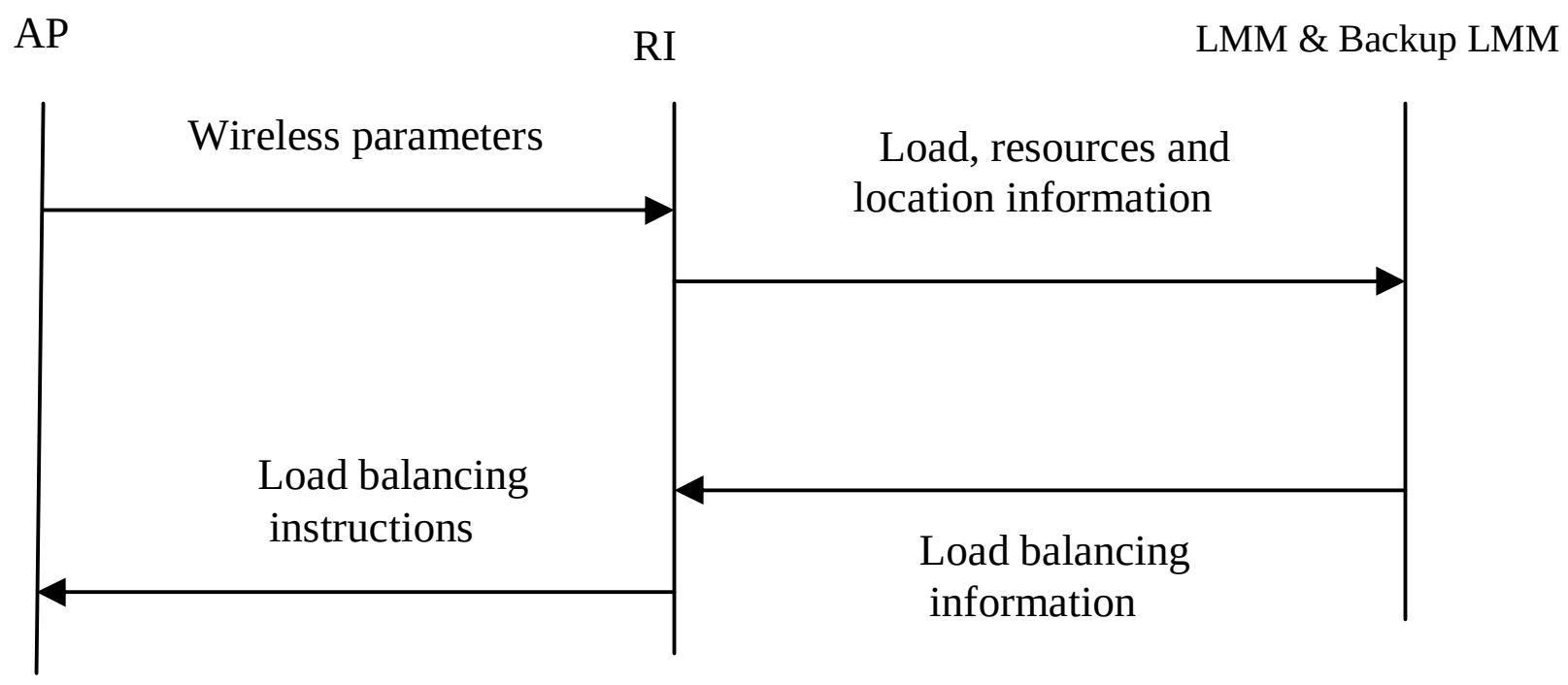

Figure 2. Signalling flow type-1

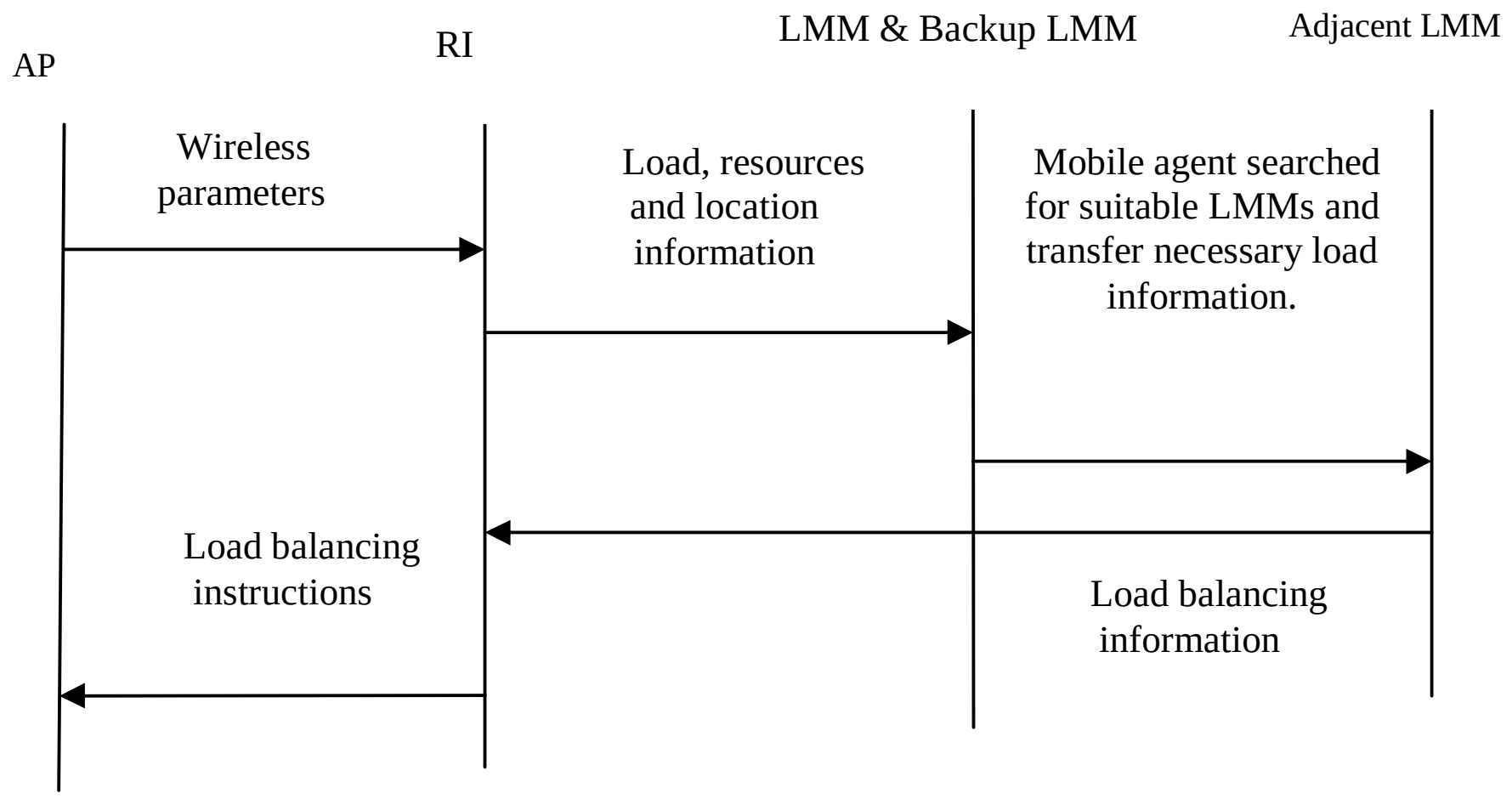

Figure 3. Signalling flow type-2

When an MS is on the border of basic grids, the mobile agent will allocate resources for the MS by consulting with the neighbouring LMMs, and the signalling flow is shown in Fig 3.

When an LMM breaks down, the backup LMM can take over immediately and the resulting signaling flow is shown in Fig 4.

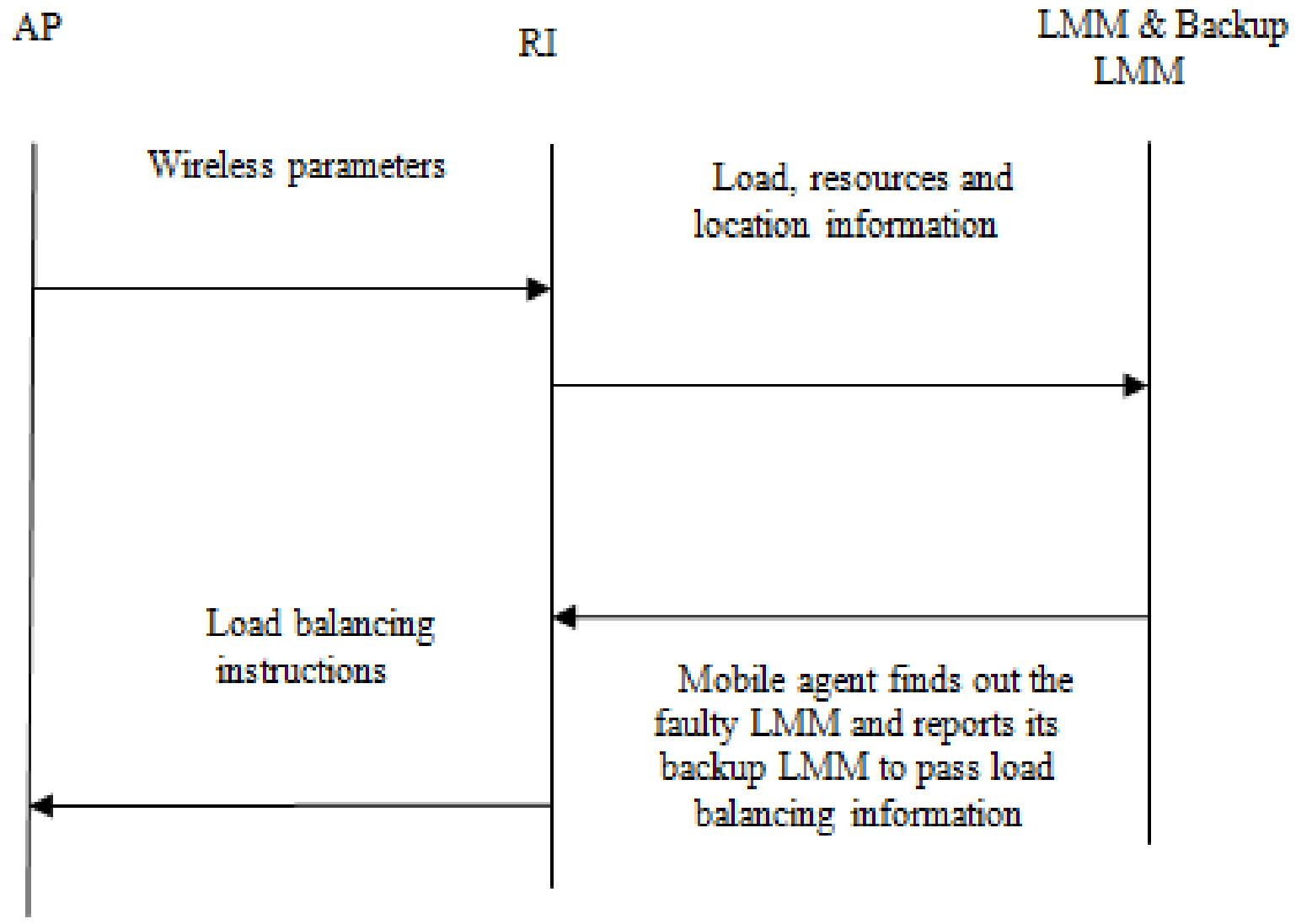

Figure 4. Signalling flow type-3

## 4. EVALUATION OF THE PROPOSED ARCHITECTURE

To evaluate the proposed architecture we have calculated the integrated reliability, signal overhead and accumulated time. Since these are significant parameters for any design, we have shown comprehensive analysis on them.

### 4.1. Signalling Overhead of the Proposed Architecture

Signalling overhead can be divided as periodic and non-periodic [11]. Periodic signalling overhead is used to backup the information. And non-periodic signalling overhead provides load information to the BB. Here we have studied both periodic and non-periodic signalling overhead.

#### 4.1.1 Periodic Signalling Overhead

For our analysis we use the following notations.

T= Time of transferring load among LMMUs.
$a_i$ = The signalling overhead of transferring load from the RI located in the system type i to the LMM.
$A_i$ = The number of APs in system type i .
d = Signaling overhead of transferring load information between the main LMM and the backup LMM.
Here, each of the signaling overheads is determined in unit time.
The signaling overhead of RIs communicating to LMMs is:

$$O_{11} = \frac{1}{T}\sum_{i=1}^{3} a_i . A_i \tag{1}$$

The signaling overhead of the main LMM communicating to the backup LMMs is:

$$O_{12} = \frac{1}{T} 2 * d \tag{2}$$

So, the total signaling overhead of transferring load information between LMMs is:

$$O_P = O_{11} + O_{12} = \frac{1}{T}\left[\sum_{i=1}^{3} a_i . A_i + 2 * d\right] \tag{3}$$

Let, $a_1=a_2=a_3$ then

$$O_P = \frac{1}{T}\left[a_1 \sum_{i=1}^{3} A_i + 2 * d\right] \tag{4}$$

#### 4.1.2 Non-Periodic Signalling Overhead

Load information transfers on the BB (which store load information for a group of LMMs and where BB resides in one of the LMM) are according to the following rules:

a) When the load state of the LMM changes then the LMM transfers load information to the BB. From the BB, the LMM can find load information of other LMMs.
b) When the main BB receives load information from the LMM then the BB immediately transfers load information to backup BBs.

Measurement of load information depends on two threshold values $k_{i1}$ and $k_{i2}$, where $k_{i1}$ is the lower threshold and $k_{i2}$ is the upper threshold. If load of any cell is less than or equal to $k_{i1}$ then the cell is considered as being in an under loaded state. If the load of any cell is greater than or equal to $k_{i2}$ then the cell is considered as being in an over loaded state. If the load of any cell is greater than $k_{i1}$ and less than $k_{i2}$ then the cell is considered as being in a balanced state.
Here we assume that the arrival and service processes are Poisson process and the queuing distribution is M/M/m. So, each cell is an independent M/M/m (m) queue.

According to the queuing theory [14], the state probabilities of system type *i* is:

$$P_{i0} = \left[ \sum_{j=0}^{m_j} \frac{(\lambda_i / \mu_i)^j}{j!} \right]^{-1} \quad (5)$$

$$P_{ik} = \frac{(\lambda_i / \mu_i)^k}{k!} P_{i0} \quad (6)$$

At time T, the probability of a cell in system type i changing from an under-loaded state to a balanced state is:

$$\Pr_i(U \to B) = P_{ik_{i1}-1} \lambda_i T \left[ 1 - (k_{i1} - 1) \mu_i T \right] \quad (7)$$

In time T, the probability that a cell in system type i changing from a balanced state to an overloaded state is:

$$\Pr_i(B \to O) = P_{ik_{i2}-1} \lambda_i T \left[ 1 - (k_{i2} - 1) \mu_i T \right] \quad (8)$$

In time T, the probability that a cell in system type i will change from an overloaded state to a balanced state is:

$$\Pr_i(O \to B) = P_{ik_{i2}+1} (k_{i2} + 1) \mu_i T (1 - \lambda_i T) \quad (9)$$

In time T, the probability that a cell in system type i will change from a balanced state to an under-loaded state is:

$$\Pr_i(B \to U) = P_{ik_{i1}+1} (k_{i1} + 1) \mu_i T (1 - \lambda_i T) \quad (10)$$

In time T, the probability that the load state of a cell changes in system type i is:

$$\Pr_i = \Pr_i(U \to B) + \Pr_i(B \to O) + \Pr_i(O \to B) + \Pr_i(B \to U) \quad (11)$$

In time T, the probability that BB j transfers load information to backup BBs is:

$$\Pr_j = 1 - \prod_{i=1}^{3} (1 - \Pr_i)^{A_{ij}} \quad (12)$$

The signalling overhead of transferring load information among the BBs is:

$$O_{NP} = \frac{1}{T} \left[ a_i . \sum_{i=1}^{3} (A_i . \Pr_i) + d . \sum_{j=1}^{2} \Pr_j \right] \quad (13)$$

## 4.2. Accumulated Processing Time of the Proposed Architecture

To derive a mathematical model for accumulated processing time, the following indicators are used.
$T_P$= Total Processing Time
$t_1$ = Time requires RI to collect and store load information (For simplicity, $t_1$ considered as 0.005 ms).
$d_{RL}$= Distance between RI and LMM
$d_{LL}$= Distance between LMM and backup LMM

$S_{RL}$ = Propagation speed between RI and LMM
$S_{LL}$= Propagation speed between LMM and backup LMM
$W_{LMM}$= Queuing delay at LMM
$T_{LMM}$= Average processing delay at LMM
$\rho$= Server Utilization Factor
$\mu$= Average service time at LMM
Propagation delay between RI and LMM

$$t_2 = \frac{d_{RL}}{s_{RL}} \tag{14}$$

Average waiting time at LMM is:

$$W_{LMM} = \frac{\rho}{\mu(1-\rho)} \tag{15}$$

Average processing delay at LMM is:

$$T_{LMM} = \frac{1}{\mu(1-\rho)} \tag{16}$$

So, total queuing delay and processing delay at LMM is:

$$t_3 = W_{LMM} + T_{LMM} \tag{17}$$

Propagation delay between LMM and backup LMM is:

$$t_4 = \frac{d_{LL}}{s_{LL}} \tag{18}$$

$W_{BLMM1}$ = Queuing delay at backup $LMM_{1.}$
$T_{BLMM1}$= Average processing delay at backup $LMM_{1.}$
Average waiting time at backup $LMM_1$ is:

$$W_{BLMM1} = \frac{\rho}{\mu(1-\rho)} \tag{19}$$

Average processing delay is at backup $LMM_1$

$$T_{BLMM1} = \frac{1}{\mu(1-\rho)} \tag{20}$$

Total queuing delay and processing delay at backup $LMM_1$ is:

$$t_5 = W_{BLMM1} + T_{BLMM1} \tag{21}$$

$W_{BLMM2}$ = Queuing delay at backup $LMM_{2.}$
$T_{BLMM2}$= Average processing delay at backup $LMM_{2.}$
Average waiting time at backup $LMM_2$ is

$$W_{BLMM2} = \frac{\rho}{\mu(1-\rho)} \tag{22}$$

Average processing delay at backup $LMM_2$ is:

$$T_{BLMM2} = \frac{1}{\mu(1-\rho)} \tag{23}$$

So, total queuing delay and processing delay at backup $LMM_2$ is:

$$t_6 = W_{BLMM2} + T_{BLMM2} \tag{24}$$

Total Processing Time is $T_P = t_1 + t_2 + t_3 + t_4 + t_5 + t_6$

$$= t_1 + \frac{d_{RL}}{S_{RL}} + \frac{\rho}{\mu(1-\rho)} + \frac{1}{\mu(1-\rho)} + \frac{d_{LL}}{S_{LL}} + \frac{\rho}{\mu(1-\rho)} + \frac{1}{\mu(1-\rho)} + \frac{\rho}{\mu(1-\rho)} + \frac{1}{\mu(1-\rho)}$$

$$= t_1 + \frac{d_{RL}}{S_{RL}} + 3\left(\frac{\rho+1}{\mu(1-\rho)}\right) + \frac{d_{LL}}{S_{LL}} \quad (25)$$

## 4.3. Integrated Reliability of the Proposed Architecture

When the number of RIs is large, the breakdown of an LMM or the junction line between an RI and the LMM has little influence on the total traffic of the system. Thus the breakdown of an RI or the junction line between an RI and the LMM can be neglected.

To derive a mathematical model for Integrated Reliability, the following indicators are used.
$R_{LMM}$= The reliability of LMM.
$R_C$= The reliability of junction lines between the base LMM and the backup LMM.
$c_j$= The traffic intensity between LMMs
$b_{ij}$= The traffic intensity between $RI_i$ and $LMM_j$.
n = The total number of LMMs.
n' = The maximum number of junction lines = n/3 + n %3 + 1
The probability that the LMM is in normal use is as follows, considering base LMM and it's two backup LMMs

$$p_{LMM} = 1 - (1 - R_{LMM})^n \quad (26)$$

Considering two junction lines between $LMM_j$ (j=1, 2, 3… … … … … n') and the two backup LMMs, the probability that the junction line is out of fault is,

$$P_C = 1 - (1 - R_C)^2 \quad (27)$$

The probability that the whole system is out of fault is

$$P_0 = P_{LMM} .P_C^{n'} .R_{LMM}^n \quad (28)$$

In that case, the traffic of the whole system does not lose,

$$L_0 = 0 \quad (29)$$

The probability that ($K_1$ =1, 2, 3…… … n'), junction lines between LMMs are out of action is,

$$P_1 = P_{LMM} \cdot \frac{n'!}{K_1!(n'-K_1)!}(1-P_C)^{K_1} .P_C^{\,n'-K_1} .R_{LMM}^n \quad (30)$$

In that case, the traffic on the $K_1$ junction lines loses, while all the traffic between LMM's and RI's does not lose. Thus the whole system traffic loss is,

$$L_1 = \sum_{j=1}^{K} c_j \quad (31)$$

The probability that $K_2$= (1, 2, 3… … … … … n), which demonstrates that LMM's are out of action, is,

$$P_2 = (1-P_{LMM}) P_C^{n'} \frac{n!}{K_2!(n-K_2)!}(1-R_{LMM})^{K_2} .R_{LMM}^{\,n-k_2} \quad (32)$$

Then, there is the need to calculate the value of another two factors: how the traffic between all the LMMs including losses irrespective of junction lines being out of action or not, and the likelihood of traffic between the $K_2$ broken LMM's turning into losses. Thus the loss of the whole system is,

$$L_2 = \sum_{j=1}^{n'} c_j + \sum_{j=1}^{k-2} \sum_{i=1}^{3} b_{ij} \tag{33}$$

The total traffic intensity of the system is

$$B = \sum_{j=1}^{n'} c_j + \sum_{j=1}^{n} \sum_{i=1}^{3} b_{ij} \tag{34}$$

The integrated reliability,

$$R = 1 - \frac{1}{B}[P_0.L_0 + P_1.L_1 + P_2.L_2] \tag{35}$$

## 5. Simulation Result

Periodic signalling overhead, non-periodic signalling overhead, processing time and integrated reliability are simulated for study and comparison purpose. The MATLAB simulator is used to compare the proposed SDA with that of the exiting HSCA [11], which is the latest flourishing architecture of heterogeneous wireless networks over SCSDA [4]. As the HSCA provides better signalling overhead and reliability than SCSDA, we consider the HSCA as our benchmark for this comparative study.

### 5.1. Simulation Setup and Scenario

For a simulation purpose and for simplicity, we consider that all three types of APs (UMTS, WiMax and WLAN) are present in each basic grid, and the number of APs of similar network types is equal in each basic grid. It is assumed that the network structures are fully overlayed. Initially subscribers or mobile nodes are randomly distributed over all of the three types of wireless networks and we consider their direction of movement arbitrarily.

The following parameters with associated values are used in our simulation study. The number of AP's in UMTS access network ($A_1$) is 600, the number of AP's in WiMax access network ($A_2$) is 900 and the number of AP's in WLAN access network ($A_3$) is 600. The time of transferring load among LMMUs *is T=0.1s.* We consider, the cell capacity of UMTS cellular system ($m_1$) is 60, WiMax system ($m_2$) is 20 and the cell capacity of WLAN system ($m_3$) is 80.

To determine the integrated reliability, initially we consider only 1 junction line among LMMs ($K_1$) are out of action and only 1 LMMs ($K_2$) are out action. For simulation study, we assume the reliability of each LMM ($R_{LMM}$) is 0.92, the reliability of junction lines between the base LMM and the backup LMM ($R_C$) is 0.97.

To simulate the signaling overhead of the proposed architecture, we initially consider the signaling overhead of transferring load from the RI (located in the UMTS system) to the LMM ($a_1$) is 1. Here, we also initially consider similar signaling overhead of transferring load from the RI to the LMM in other two cellular system (WiMax and WLAN) that is $a_1=a_2=a_3=1$. The signaling overhead of transferring load information between the main LMM and the backup LMM ($d$) is considered as 1.

To simulate accumulated processing time, we consider 0 .005ms as the required time ($t_1$) of RI to collect and store load information. For simplicity, both the distance from RI to LMM ($d_{RL}$) and the distance from LMM to backup LMM ($d_{LL}$) are considered as 500 meters. For simulation purpose, the propagation speed between RI and LMM ($S_{RL}$) is assumed as 0.5Mbps, which is equivalent to the propagation speed between LMM and backup LMM ($S_{LL}$).

## 5.2. Simulation Study

The simulation result of the periodic signalling overhead is shown in Figure 5. Here we consider discreet number of LMMs as (1, 50, 100, 150, 200, 250 and 300). We can see that the periodic signalling overhead of our proposed system (SDA) remains constant, whereas the periodic signalling overhead of the HSCA increases with the increment of RAs, and the overhead of HSCA is always higher than the overhead of the SDA. We can see in [4] that the periodic signalling overhead of the SDA is not increased with the increase of LMMs because one LMM is connecting only and at most two neighbouring backup LMMs for any network size. On the other hand by increasing RA in HSCA, the overhead always increases.

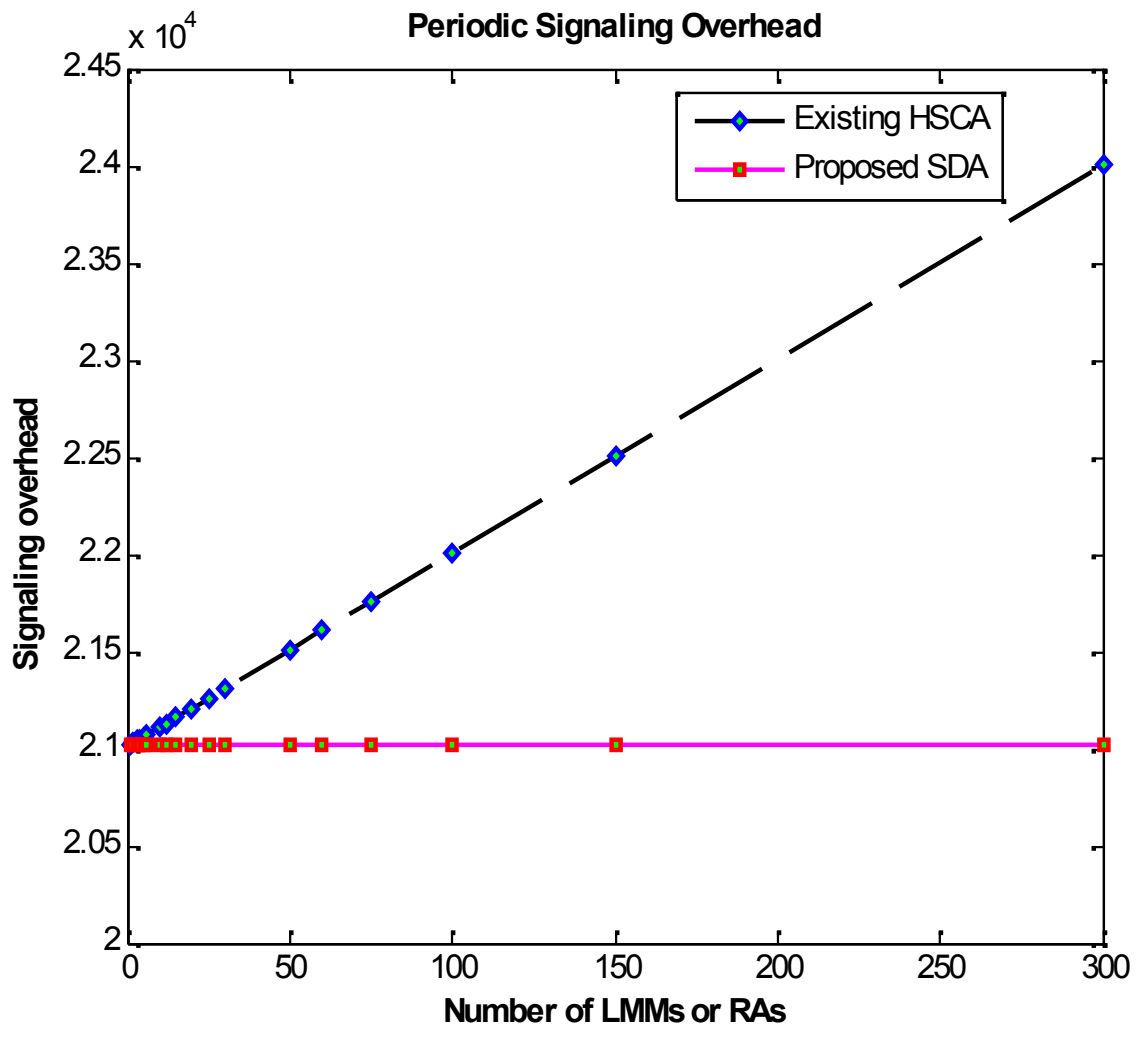

Figure 5. Periodic signalling overhead

The signalling overhead for a non-periodic signal is shown in Figure 6. Here, the non-periodic signalling overhead remains fixed in the proposed SDA with the increment of LMMs. On the other hand, the non-periodic signalling overhead of the HSCA is increased exponentially with the increment of RAs. As there is no centralized LMM in the proposed SDA, increment of LMMs don't have any impact on the overall non-periodic signalling overhead, which we have analyzed in [13].

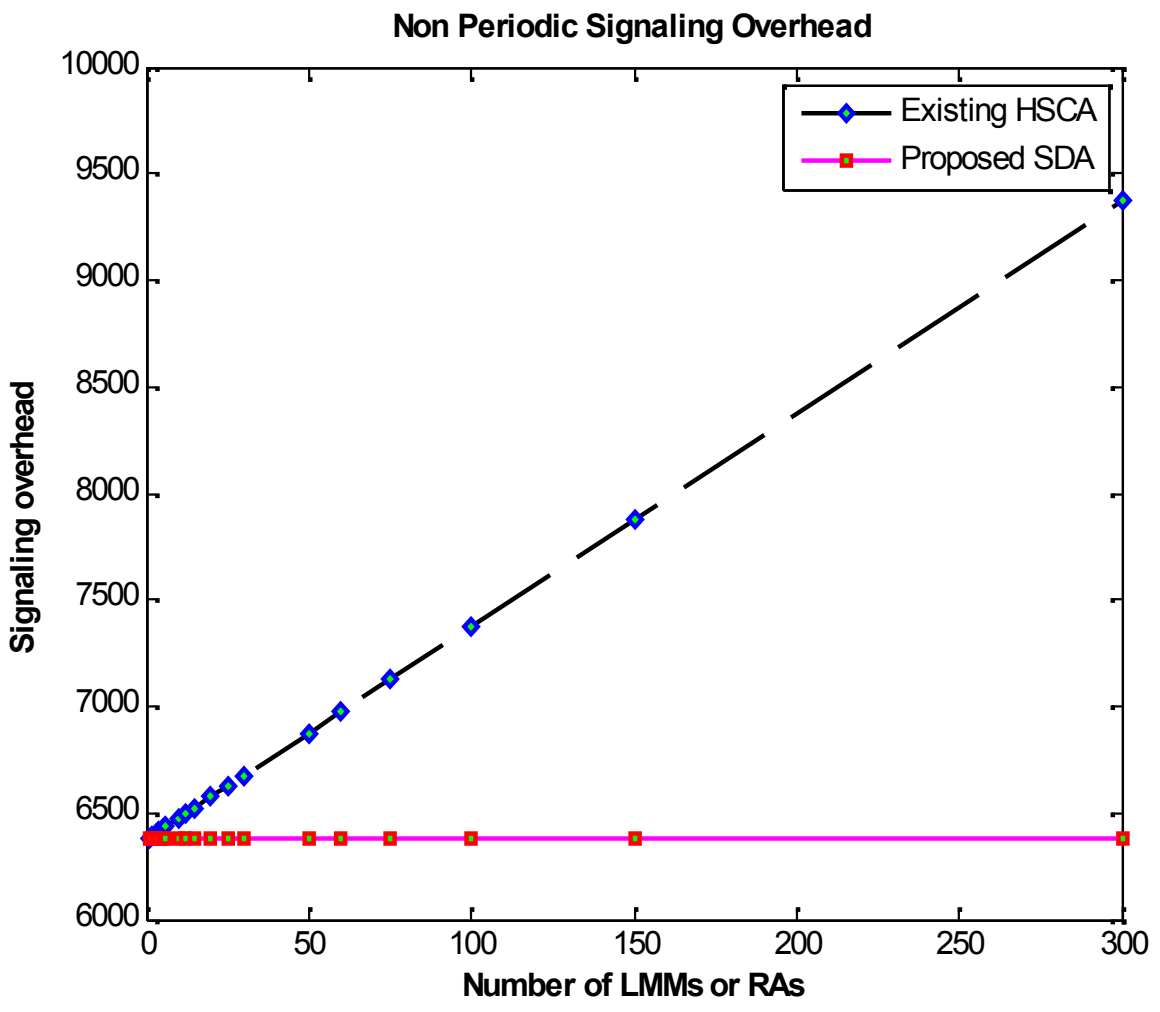

Figure 6. Non-periodic signalling overhead

Figure 7 shows the overall processing time of the proposed SDA architecture with that of the existing HSCA architecture. The processing time of the HSCA according to [36] is much greater than the processing time of the SDA. In both models, the processing time increases linearly with the increment of the traffic arrival rate. The processing time difference between the HSCA and the SDA also increases in linear fashion.

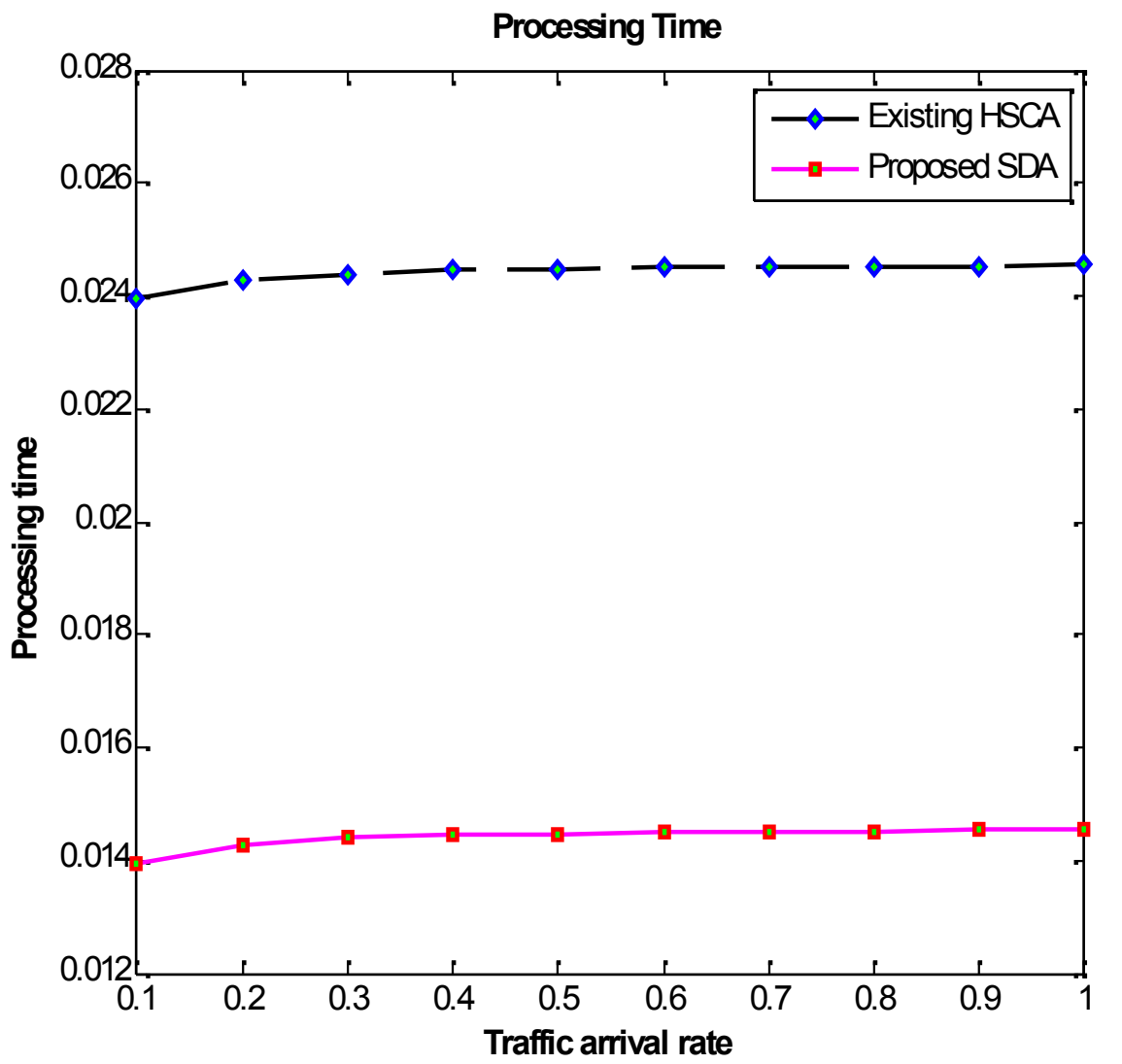

Figure 7. Overall processing time

Figure 8 shows the integrated reliability of the proposed architecture (SDA) and the integrated reliability of existing HSCA architecture. The integrated reliability is increased with the increment of the number of LMMs (RAs) in both of the architecture, but the increment rate of SDA is more faster than the increment rate of HSCA.

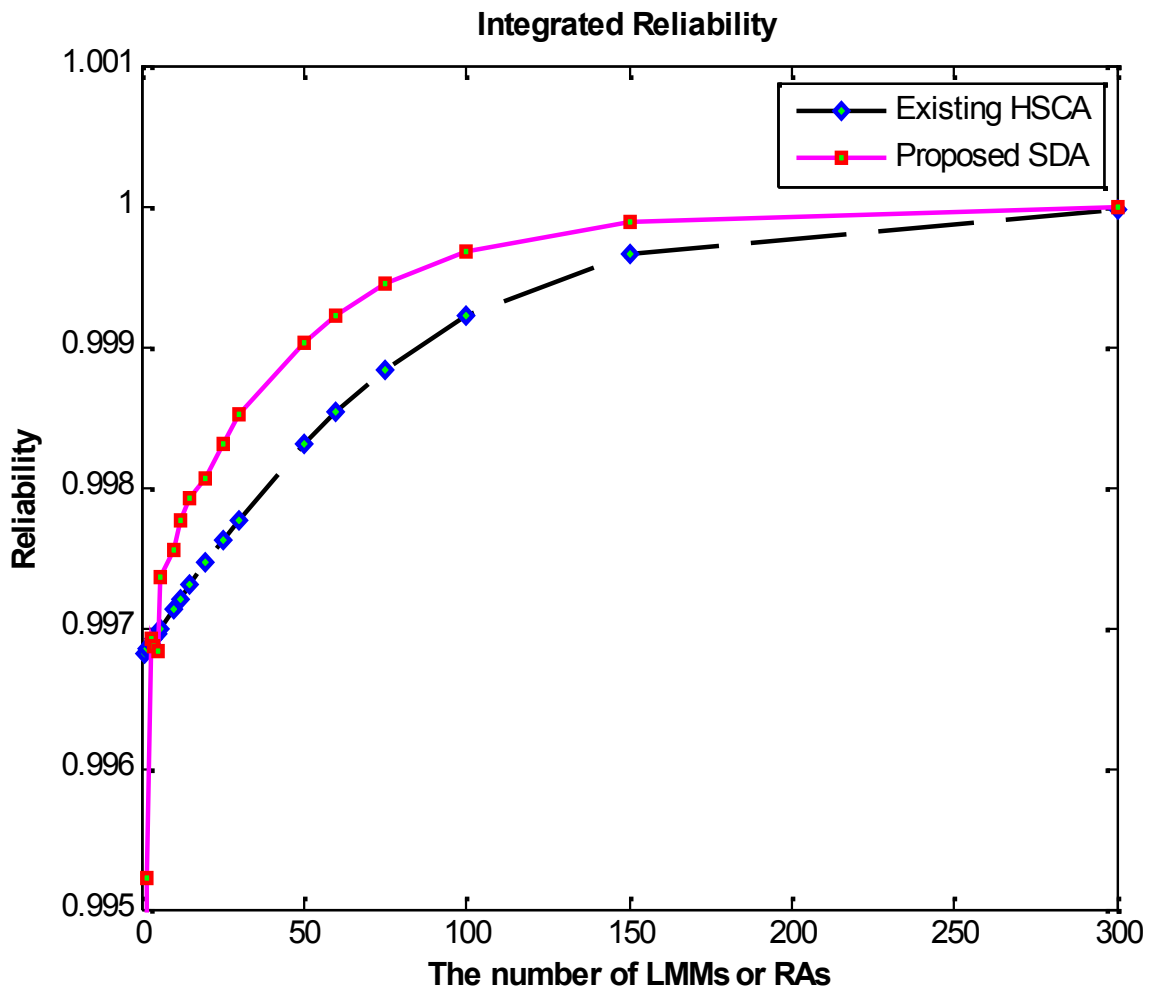

Figure 8. Integrated reliability

## 6. Conclusions

The proposed architecture is semi-distributed in nature. We have first proposed SDA architecture in [15], and later, details of the design and assessment of various parameters are articulated here. From a simulation study, it can be easily known that the SDA performs better by increasing reliability and minimizing overhead and time. In hierarchal semi-centralized architecture [11] there exist only two ISs for the whole network. If they fail to perform then the total network operation will be hampered. But the SDA is free from one 'single' point of failure. And it works better for two challenging parameters of distributed system. Obviously, the proposed design introduces an efficient architecture in heterogeneous wireless network.

## Appendix

The authors of the reference [11] presented a 'load balancing' architecture named HSCA, where they introduced a resource management unit (RMU), which consists of Resource Statistics (RS), a Resource Allotter (RA), an Information Server (IS) and a backup IS. The RS calculates the load and resources of its jurisdiction cell, the execution of which requires a certain period of time ($t_1$). Then the RS transfers the load resources and location information to the RA, which requires propagation delay between the RS to the RA, queuing delay and processing delay within RAs. Next, the RA transfers the load resources and location information to the IS, which requires propagation delay between the RA to the IS, queuing delay and processing delay within ISs. Similarly the propagation delay, queuing delay and processing delay within the backup ISs exists when the main IS transfers load and resource information to the backup IS. Thus the total processing time is:

$$T_p = t_1 + \frac{d_{RR}}{S_{RR}} + \frac{d_{RIS}}{S_{RIS}} + \frac{d_{IBI}}{S_{IBI}} + \frac{1}{\mu(1-\rho_{RA})}(\rho_{RA}+1) + \frac{2}{\mu(1-\rho_{IS})}(\rho_{IS}+1) \quad (36)$$

## Authors

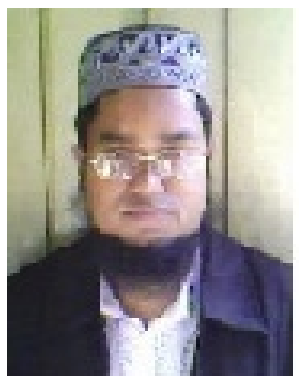

Md Golam Rabiul Alam received the Bachelor Degree in Computer Science & Engineering from Khulna University in 2002 and the Master Degree in Information Technology from University of Dhaka in 2011. He is now an Assistant Professor of Department of Computer Science and Engineering in International Islamic University Chittagong (IIUC). His current research interests include the Heterogeneous Wireless Networks, Data Mining and Information Security. He is a co-author of about thirty research papers and a text book.

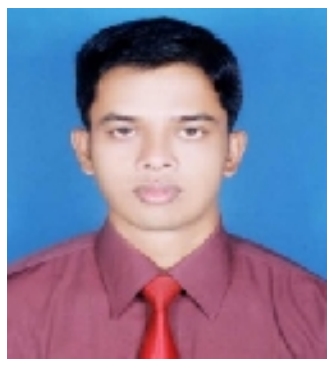

Chayan Biswas received the Master Degree in Information Technology from University of Dhaka in 2011. He is now a Senior Officer (IT) of Bangladesh Development Bank Limited (BDBL). His current research interests include the Load Balancing technology for Wireless Heterogeneous Networks and Wireless Security.